\documentclass[twocolumn]{aastex6}
\shorttitle{Chromospheric heating by acoustic waves}
\shortauthors{M. Sobotka et al.}

\begin{document}

\title{Chromospheric heating by acoustic waves compared to radiative cooling}

\author{M. Sobotka\altaffilmark{1}, P. Heinzel\altaffilmark{1}, M. \v{S}vanda\altaffilmark{1,2},
J. Jur\v{c}\'ak\altaffilmark{1}, D. del Moro\altaffilmark{3}, and F. Berrilli\altaffilmark{3}}

\altaffiltext{1}{Astronomical Institute, Academy of Sciences of the Czech Republic (v.v.i.), Fri\v{c}ova 298, 25165 Ond\v{r}ejov, Czech Republic}
\altaffiltext{2}{Astronomical Institute, Charles University in Prague, Faculty of Mathematics and Physics, V Hole\v{s}ovi\v{c}k\'ach 2, 18000 Prague 8, Czech Republic}
\altaffiltext{3}{Department of Physics, University of Roma Tor Vergata, Via della Ricerca Scientifica 1, 00133 Rome, Italy}

\begin{abstract}
Acoustic and magnetoacoustic waves are among the possible candidate mechanisms that heat the upper layers of solar atmosphere. A weak chromospheric plage near a large solar pore NOAA 11005 was observed on October 15, 2008 in the lines \ion{Fe}{1} 617.3~nm and \ion{Ca}{2} 853.2~nm with the Interferometric Bidimemsional Spectrometer ({\em IBIS}) attached to the Dunn Solar Telescope. Analyzing the \ion{Ca}{2} observations with spatial and temporal resolutions of 0\farcs 4 and 52 s, the energy deposited by acoustic waves is compared with that released by radiative losses. The deposited acoustic flux is estimated from power spectra of Doppler oscillations measured in the \ion{Ca}{2} line core. The radiative losses are calculated using a grid of seven 1D hydrostatic semi-empirical model atmospheres. The comparison shows that the spatial correlation of maps of radiative losses and acoustic flux is 72 \%. In quiet chromosphere, the contribution of acoustic energy flux to radiative losses is small, only of about 15~\%. In active areas with photospheric magnetic field strength between 300 G and 1300 G and inclination of 20\degr--60\degr, the contribution increases from 23~\% (chromospheric network) to 54~\% (a plage). However, these values have to be considered as lower limits and it might be possible that the acoustic energy flux is the main contributor to the heating of bright chromospheric network and plages.
\end{abstract}


\section{Introduction}
\label{Sect:1}

The solar chromosphere looses energy mostly by radiation in lines of \ion{Ca}{2} and \ion{Mg}{2}, in Lyman $\alpha$ and the H$^-$ continuum \citep{Vernazza81,Stix04}. The total radiative losses integrated over the height of the chromosphere are of the order of 4300 W m$^{-1}$ in the quiet Sun \citep{Avrett81}. In active regions, they are higher by a factor of 2 to 4 \citep{Withbroe77}. These losses have to be balanced by an energy input. Several candidate mechanisms to heat the solar chromosphere have been proposed and studied \citep[see][for a review]{Jess15}. They can be divided into two main classes: magnetic reconnection and waves.

In the first case, it is suggested that regular reconfigurations of the embedded magnetic field lines will produce extreme localized heating through the release of magnetic energy \citep{PriestSchrijver99}. Rapidly occurring small-scale flare events, ``nanoflares'', with individual energies $\sim 10^{17}$ J, may occur with such regularity in the solar atmosphere that they can provide the continual source of heat required to maintain the elevated temperatures in the corona \citep{Parker88} and also in the chromosphere \citep{Testa14}. Unfortunately, the small spatial sizes of such events places them within or below the noise threshold of current observations \citep{Terzo11}.

In the second case, waves, generated near the solar surface through turbulent motions of the convective plasma, propagate upwards and dissipate a considerable part of their energy in the chromosphere. In the magnetized solar atmosphere with local magnetic field often exceeding 1 kG, oscillatory modes become highly modified, producing anisotropic waves that can be modelled using magnetohydrodynamic (MHD) approximations. There exist three types of waves: Alfv\'en, fast and slow magnetoacoustic waves \citep{Khomenko09,KhomenkoCally12}. Low-frequency MHD waves may represent a significant source of energy related to so-called {\em magnetic portals} \citep{Jefferies06}. Although these waves are generally not allowed to propagate higher into the atmosphere, because their frequency does not exceed the expected photospheric cutoff frequency $\nu_{\rm c} = 5.2$~mHz \citep{BelLeroy77}, in regions where the photospheric magnetic field is largely inclined with respect to the gravity vector the cutoff frequency can be lowered by means of the ramp effect \citep{Stangalini11}. This allows the propagation of waves with frequencies far below 5.2 mHz, which would otherwise be trapped in the photosphere, into the upper atmosphere.

The atmospheric density significantly decreases from the photosphere to the chromosphere, and as a result radiative transition rates generally dominate over collisional ones. This makes the chromosphere a non-LTE environment, resulting in the need for full radiative transfer modelling of all simulated processes. The chromosphere has been described using semi-empirical 1D hydrostatic models, particularly by \citet[][VAL models]{Vernazza81}, spanning from an intranetwork (VAL A,~B) over an average quiet Sun (VAL~C) and average network (VAL~D) to bright network elements (VAL E,~F). This set of six models was complemented and extended by \citet[][FAL models]{Fontenla93,Fontenla99,Fontenla06} who, in addition to quiet Sun models (FAL A--F), included plages (FAL~H), faculae (FAL~P), bright faculae (FAL~Q), and also sunspots (FAL R,~S).

Chromospheric heating by dissipation of (magneto)acoustic waves is generally time-dependent. \citet{Carlsson95,Carlsson97} calculated the propagation of acoustic waves using time-dependent hydrodynamic models including non-LTE radiative transfer. They found that short intervals of very high temperature are caused by acoustic shocks but the average chromospheric temperature continues to decrease with height analogously to the photosphere. Nevertheless, from the calculated emission in the \ion{Ca}{2} H, K lines averaged in time they deduced a ``semi-empirical'' mean temperature that resembles the 1D hydrostatic models similar to VAL. This is an articact of a non-linear time averaging of UV line intensities. However, this would conflict with the observation that UV lines originating in the solar chromosphere are in emission everywhere and all of the time \citep{Carlssonetal97,Curdt98}. Even multi-dimensional non-LTE models are not entirely representative of the observed chromospheric structures due to high computational requirements and inevitable simplifications \citep{CarlLeen12}. For these reasons, the use of 1D hydrostatic models that provide time-averaged atmospheric parameters could still be viable.

Recently, the signatures of (magneto)acoustic waves can be observed with high spatial resolution. The work of \citet{Bello09,Bello10} strengthened the support for atmospheric heating through magnetoacoustic wave dissipation. For example, \citet{Sobotka13} claimed that the chromosphere above a light bridge, located inside a large pore, can be heated thanks to the propagation of low-frequency ($< 5.2$ mHz) waves. In the present work we attempt to estimate observationally the contribution of acoustic waves to the chromospheric heating of quiet and plage regions, comparing the deposited acoustic energy flux with the radiative cooling.

\section{Observations}
\label{Sect:2}

\begin{figure}[t]
       \centering
\includegraphics[width=0.46\textwidth]{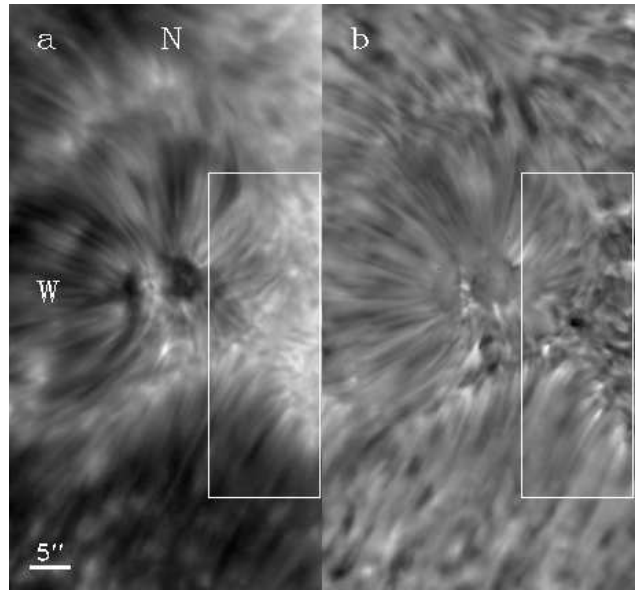}
\caption{A snapshot at 17:10 UT of the full FOV (38\arcsec $\times$ 71\farcs 5) with the pore NOAA 11005. North is at the top, west to the left. (a) \ion{Ca}{2} line center intensity displayed in logarithmic scale for better visualization, (b) Doppler map at $h \sim 1000$~km. The velocities range from $-4.5$ km~s$^{-1}$ (black, toward the observer) to 5.9 km~s$^{-1}$ (white, away from the observer). The region of interest is outlined by a white rectangle. Panel (a) is also available as an animation.}
       \label{Fig:fullFOV}
   \end{figure}

A large isolated solar pore in the active region NOAA 11005 was observed on October 15, 2008 from 16:34 to 17:44 UT with the Interferometric Bidimensional Spectrometer \citep[{\em IBIS},][]{Cavallini06} at the Dunn Solar Telescope ({\em DST}), Sunspot, NM, equipped with an adaptive optics system \citep{Rimmele04}.
The bipolar active region NOAA 11005 lived from October 11 to 16, 2008. The pore was present in the leading part of the region. During the observations, it was  located at 25.2~N and 10.0~W (heliocentric angle $\vartheta = 23$\degr, $\mu = \cos(\vartheta) = 0.92$) and decayed slowly. The pore was surrounded by a superpenumbra \citep{Sobotka13}. A weak chromospheric plage with the same magnetic polarity as the pore was located at the eastern edge of the field of view (FOV) shown in Fig.~\ref{Fig:fullFOV}.

%
%

We use the {\em IBIS} dataset described by \citet{Sobotka12,Sobotka13}. Each of the 80 data sequences contains a full Stokes ($I, Q, U, V$) 21-point scan of the \ion{Fe}{1} 617.33~nm line (wavelength spacing 2~pm) and a 21-point $I$-scan of the \ion{Ca}{2} 854.2 nm line (wavelength spacing 6~pm). At each wavelength in the \ion{Fe}{1} line, six modulation states $I \pm [Q, U, V]$ were acquired. Each sequence therefore consists of $21 \times  6$ (\ion{Fe}{1}) $ +\ 21$ (\ion{Ca}{2}) = 147 narrowband images. With an exposure time of 80~ms for a single image, one sequence took 52~s, which is the time resolution. From this, the $I$-profile of the \ion{Ca}{2} line was scanned in 6.4~s. The size and pixel scale of the narrowband images is $512 \times 512$ pixel and 0\farcs 167, respectively. As a consequence of the {\em IBIS} polarimetric setup, the FOV was reduced to $228 \times 428$ pixels, that is, 38\arcsec $\times$ 71\farcs 5, with the pore located in the center. Synchronously with the narrowband images, broadband (WL) frames at $\lambda = 621.3 \pm 5$~nm were acquired with an equal exposure time of 80~ms.

From the analysis of spectropolarimetric observations in the \ion{Fe}{1} 617.33~nm line of the same dataset \citep{Sobotka12}, the spatial resolution was estimated to be of 0\farcs 4, the spectral stray-light contamination of 1--2~\%, and the spatial stray light of 15~\%. The instrumental characteristics of {\em IBIS} for the \ion{Ca}{2} 854.2 nm line were described by \citet{Cauzzi08}. At 854~nm, the {\it FWHM} of the instrumental spectral transmission is 4.4~pm \citep{Reardon08} and the spectral stray-light contamination is of 1.5~\%.

\section{Data processing}
\label{Sect:3}

The processing of data is described in detail by \citet{Sobotka12,Sobotka13}. The narrowband images were corrected for motion and distortion, using as a reference a series of WL frames reconstructed by the multi-frame blind deconvolution method \citep[MFBD,][]{Noort06}. Then, the standard reduction pipeline \citep{Viticchie10} was applied to the narrowband data to correct for the instrumental blueshift \citep{Cavallini06} and the instrument- and telescope-induced polarizations.
In this paper we focus on observations in the infrared \ion{Ca}{2} 854.2 nm line. According to \citet{Cauzzi08}, the wings 50--70~pm far from the line center come from the middle photosphere 200--300~km above the unity optical depth in the 500 nm continuum ($\tau_{500} = 1$), while the line core is formed in the middle chromosphere between 900 and 1500~km.

The \ion{Ca}{2} 854.2 nm core shows significant Doppler shifts caused by chromospheric oscillations. Line-of-sight (LOS) velocity maps were measured (1) in the inner wings of the core at $\pm 18$ pm that are formed approximately at 1000~km, using the {\em double-slit} method \citep{Garcia10}, and (2) in the line center formed at about 1400--1500~km, using a parabolic fit to five wavelength points. The reference zero of the Doppler velocity was defined as a time- and space average of all measurements. Measurements (1) and (2) gave practically identical results, with slightly (by 4 \%) higher amplitudes detected by (2).

\begin{figure}[t]
       \centering
\includegraphics[width=0.50\textwidth]{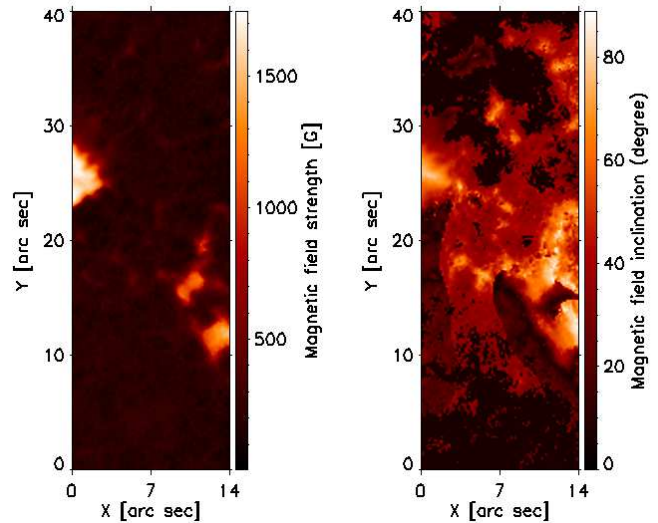}
\caption{Maps of photospheric magnetic field strength and inclination in the region of interest.}
       \label{Fig:magfield}
   \end{figure}

The magnetic field configuration in the photosphere,  necessary for the calculation of the acoustic energy flux in Sect.~\ref{Sect:5}, was retrieved using the Stokes inversion code based on response functions \citep[SIR,][]{RuizCobo92}, applied to the \ion{Fe}{1} 617.33~nm Stokes profiles. Two nodes were set for the temperature and one node for all the other parameters such as the magnetic-field vector and the LOS velocity. The microturbulent velocity was fixed at 1.3~km~s$^{-1}$, the macroturbulence at zero, the spatial stray-light contamination was set to 15 \%, and the {\em IBIS} spectral instrumental profile was included in the inversion.
The retrieved magnetic field inclination and azimuth were transformed to the local reference frame using the routines of the AZAM code \citep{Lites95} and the
azimuth ambiguity was resolved using the method by \citet{Georgoulis05}. We selected a scan at 17:10 UT in the middle of the observation to represent the magnetic field configuration. Since the Stokes $Q$ and $U$ signals were noisy in regions with weak magnetic field strength ($B < 300$~G), resulting in false values of the inclination $\theta$ near 90\degr \  (horizontal direction), a mask was applied to set $\theta > 85$\degr \  to zero in such regions. For $B > 300$~G all the inclination values were allowed. The region of interest that covered the plage to the east of the pore together with some quiet Sun area was set to $82 \times 238$ pixels, that is, 13\farcs 7 $\times$ 39\farcs 7 (Fig.~\ref{Fig:fullFOV}). The maps of photospheric magnetic field strength and inclination retrieved from the inversion of the \ion{Fe}{1} 617.33 nm Stokes profiles are shown in Fig.~\ref{Fig:magfield}.

%
%

\section{Radiative cooling}
\label{Sect:4}

The solar chromosphere is very dynamic by its nature, hosting waves and oscillations with periods typically of few minutes. However, the structures related to photospheric magnetic fields (chromospheric network, plages, superpenumbrae around spots and pores) are preserved for several hours and longer. An animation of Fig.~\ref{Fig:fullFOV}a shows the temporal evolution of the observed \ion{Ca}{2} line center intensity in the full FOV including the region of interest (outlined by a rectangle) and an intensity map averaged over the 70-minute period of observation, attached at the end of the movie. We assume that an analysis of time-integrated observables like acoustic power maps measured from the series of Dopplergrams \citep[Fig.~7 of][]{Sobotka13} and time-averaged profiles of spectral lines together with the atmospheric parameters taken from 1D semi-empirical hydrostatic models can provide us with at least a rough statistical information about an energy balance in the long-lived chromospheric structures.

The energy released by radiation from the chromosphere is characterized by net radiative cooling rates (radiative losses), which, on the other hand, indicate the amount of non-radiative heating that sustains the temperature at a given height in the atmosphere. To calculate them, we  approximate the atmosphere in the region of interest by a grid of semi-empirical models VAL A--F and FAL H, P (see Sect.~\ref{Sect:1}). Synthetic profiles of the \ion{Ca}{2} 854.2~nm line are computed for each model using non-LTE radiative-transfer code based on the MALI technique \citep{Rybicki91,Rybicki92} with standard partial frequency redistribution (PRD, angle-averaged) in the \ion{Ca}{2} H and K resonance lines. In the first step, the hydrogen version of the code (with PRD in Lyman lines) computes the ionization structure and populations of hydrogen levels, in the second step the \ion{Ca}{2} version is used with an atomic model including five levels and continuum.

%
   \begin{figure}[]
       \centering
\includegraphics[width=0.50\textwidth]{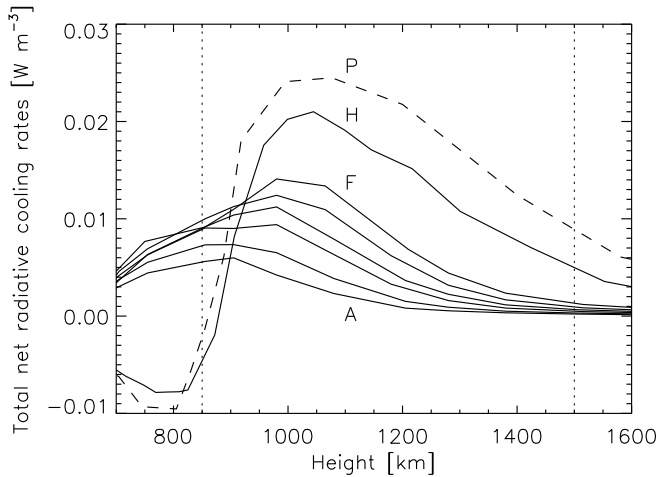}
\caption{Height distribution of total net radiative cooling rates calculated from models VAL A--H and FAL H, P. Vertical dotted lines delimit the range of integration.}
       \label{Fig:rlosses}
   \end{figure}
%

For each model, the same code is used to compute the net radiative cooling rates for the lines \ion{Ca}{2} H, K and the infrared triplet, \ion{Mg}{2} h, k and hydrogen lines, and hydrogen continua. The total net cooling rates, a sum of the contributions mentioned above, are plotted in Fig.~\ref{Fig:rlosses}. The total radiative losses are then integrated over geometrical heights in the range from 850 km to 1500 km, where the cooling rates are significant. In an approximately equal height range we measure the \ion{Ca}{2} 854.2~nm Doppler signal and get the information about the acoustic energy flux. At higher altitudes, also the hydrogen L$\alpha$ line contributes, but that region is not studied in this work. The total integrated radiative losses are listed in Table~\ref{Tab:losses}.

%
\begin{table}[b]
\caption{Total radiative losses integrated over geometrical height.}
\label{Tab:losses}
\centering
\begin{tabular}{ccc}
\hline \hline
\noalign{\smallskip}
    Model & Rad.\ losses [W m$^{-2}$] & Area occupied [\%] \\
\hline
\noalign{\smallskip}
    VAL A &  1830 & 14 \\
    VAL B &  2600 & 14 \\
    VAL C &  3630 & 21 \\
    VAL D &  4260 & 17 \\
    VAL E &  5130 & 15 \\
    VAL F &  5930 & 13 \\
    FAL H &  8220 &  6 \\
    FAL P & 10040 &  -- \\
\hline
\end{tabular}
\tablecomments{Second column includes relative areas occupied by individual models in the region of interest.}
\end{table}
%

It is well known that the computed \ion{Ca}{2} line profiles are very sensitive to expected microturbulent velocity. High microturbulence velocity smooths small emission peaks on both sides of the line core, while zero microturbulence leads to a large enhancement of them, as shown, e.g., in the \ion{Ca}{2} modeling by \citet{Uitenbroek89}. Since we do not know the realistic distribution of microturbulent velocities, which can also be related to waves and oscillations in the chromosphere, we just use the values supplied by the models. Only for the model FAL H with original microturbulent velocities higher nearly by a factor of two than the VAL~A--F and FAL~P values, resulting in an improbably broad synthetic profile, we use the microturbulent velocity distribution of FAL~P.

%
   \begin{figure}[t]
       \centering
\includegraphics[width=0.5\textwidth]{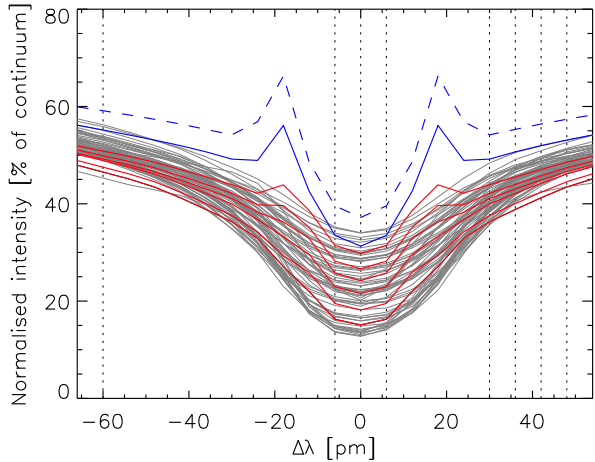}
\caption{\ion{Ca}{2} 854.2~nm line in the region of interest. Observed (gray), VAL A--F synthetic (red), and FAL H, P synthetic profiles (blue). FAL~P (dashed) does not match the observations at all. Vertical dotted lines show wavelengths used for the fit.}
       \label{Fig:profiles}
   \end{figure}
%

\begin{figure}[t]
       \centering
\includegraphics[width=0.50\textwidth]{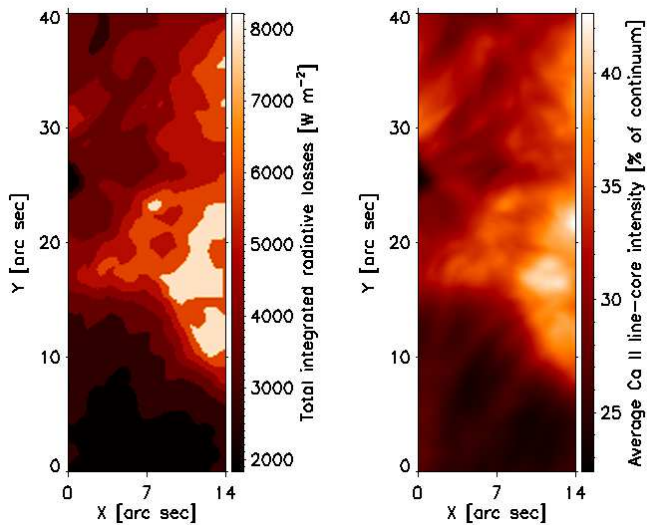}
\caption{Maps of total integrated radiative losses and time-averaged \ion{Ca}{2} 854.2~nm line center intensity.}
       \label{Fig:lossmap}
   \end{figure}

To assign appropriate models and also radiative losses to various chromospheric structures, for each pixel in the region of interest we calculate mean profiles of the \ion{Ca}{2} 854.2~nm line, averaging the observed profiles (with Doppler shifts removed) over the 70-minute period of observation. The mean profiles are normalized to the continuum intensity at $\mu = 0.92$ using a fit of quiet-Sun profiles to the atlas profile \citep{Linsky70}. The models are assigned according to the best fit of the synthetic profiles to mean observed profiles at eight wavelength points, avoiding the near-wing emission peaks that are never observed in our data set. The wavelength points were selected in such a way that the contribution to the fit from the wings is comparable with the contribution from the core of the profile. Fig.~\ref{Fig:profiles} shows the match of the mean observed profiles (for every 20-th pixel in the region) with the synthetic ones, together with the wavelengths used for the fit. The FAL~P model is not used, because its synthetic profile does not match the observations; profiles calculated for all other models represent our observations sufficiently both in the line core and wings. In this way we obtain a map of total integrated radiative losses in the region of interest, matching well with the time-averaged line center intensity distribution. Both maps are displayed in Fig.~\ref{Fig:lossmap}. Relative areas occupied by individual models in the region of interest are listed in Table~\ref{Tab:losses}.

%
%

\section{Deposited acoustic energy flux}
\label{Sect:5}

Oscillations in various areas in the FOV are studied using standard Fourier analysis of the time-series of Doppler velocity maps measured in the \ion{Ca}{2} 854.2~nm line at two different heights in the chromosphere, 1000~km and 1500~km approximately (see Sect.~\ref{Sect:3}).
Given the length of the time-series 70 minutes, the frequency resolution is 0.24~mHz. This, together with the sampling period of 52 s that sets the maximum detectable frequency to 9.6~mHz, is fine enough to investigate the low-frequency part of oscillation power spectra in the quiet-Sun and magnetic regions in the chromosphere. Power maps for nine frequency bands from 1~mHz to 9~mHz in the full FOV were calculated by \citet{Sobotka13} and six of them were displayed in their Fig.~7.

Following \citet{Bello09}, we use the observed Doppler velocities to estimate the acoustic energy flux at the two chromospheric heights in the region of interest. The method of power spectra calculation and calibration in absolute units is described by \citet{Rieutord10}. The acoustic power flux can be estimated from the equation
\begin{equation}
F_{\rm ac,tot}=\int\limits_{\nu_{\rm ac}}^{\nu_{\rm max}} \rho P_{\rm v}(\nu)
\upsilon_{\rm gr}(\nu)/T\!F(\nu)\, {\rm d}\nu,
\label{eq:acousticflux}
\end{equation}
where $\rho$ is the gas density, $P_{\rm v}$ is the spectral power density derived from the velocities, and $\upsilon_{\rm gr}$ is the group velocity of energy transport, given by
\begin{equation}
\upsilon_{\rm gr}=c_{\rm s}\sqrt{1-(\nu_{\rm ac}/\nu)^2}.
\label{eq:groupvel}
\end{equation}
In this formula, $c_{\rm s}$ is the sound speed
\begin{equation}
c_{\rm s} = \sqrt{\gamma p/\rho}
\end{equation}
and $\nu_{\rm ac}$
is the acoustic cutoff frequency given by
\begin{equation}
\nu_{\rm ac}=\frac{\gamma g \cos{\theta}}{4\pi c_{\rm s}},
\end{equation}
with  $\gamma$ being the adiabatic index equal to 5/3 for monoatomic gas, $g$ the surface gravity and $\theta$ the magnetic field inclination in the photosphere. The effect of magnetic-field inclination that reduces the acoustic cutoff frequency was analytically derived by \citet{Cally06}.

%
   \begin{figure}[t]
       \centering
\includegraphics[width=0.50\textwidth]{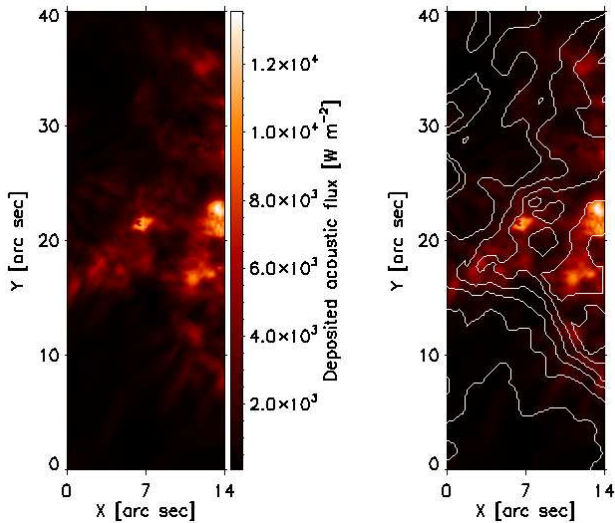}
\caption{Map of the deposited acoustic energy flux. In the right panel, the map is overlaid by contours of the total integrated radiative losses shown in Fig.~\ref{Fig:lossmap}.}
       \label{Fig:acflux}
   \end{figure}
%

The values of gas pressure $p$ and density $\rho$ at the two heights in the chromosphere are taken from the corresponding models VAL and FAL assigned to each pixel in the region of interest (Sect.~\ref{Sect:4}). The magnetic field inclination $\theta$ is obtained for each pixel form the inversion of the \ion{Fe}{1} 617.33~nm Stokes profiles (Sect.~\ref{Sect:3} and Fig.~\ref{Fig:magfield}).
The function $T\!F(\nu)$ in Equation~(\ref{eq:acousticflux}) denotes the {\em transfer function} of the solar atmosphere \citep{Bello09}. Its value equal to one means that the observed Doppler signal of all waves at a given frequency is detected throughout the solar atmosphere, whereas values smaller than one represent some loss of the signal due to the extent in height of the spectral-line contribution functions. In general, the most affected are short-period (small-scale) fluctuations. A detailed time-dependent model of the atmosphere is needed to derive it. For the estimate in this paper we did not attempt to calculate the correct $T\!F(\nu)$ and set it to one for all frequencies. However, when $T\!F(\nu) < 1$, the total acoustic energy flux may increase substantially.

It can be expected that a part of the acoustic flux energy estimated for lower layers around $h_1=1000$~km is dissipated and converted into radiation in the height range 850--1500~km (Sect.~\ref{Sect:4}). The rest of the acoustic flux, estimated for upper layers around $h_2=1500$~km, continues to propagate higher in the atmosphere. Thus, the acoustic energy flux $\Delta F_{\rm ac}$ deposited into the middle chromosphere between the heights $h_1, h_2$ can be expressed as a difference
\begin{equation}
\Delta F_{\rm ac}=F_{\rm ac,tot}(h_1) - F_{\rm ac,tot}(h_2).
\label{eq:depflux}
\end{equation}
On average, $F_{\rm ac,tot}(h_2) / F_{\rm ac,tot}(h_1) \simeq 0.1$. The resulting map of the deposited acoustic energy flux in the region of interest is displayed in Fig.~\ref{Fig:acflux}.

%
   \begin{figure}
       \centering
\includegraphics[width=0.48\textwidth]{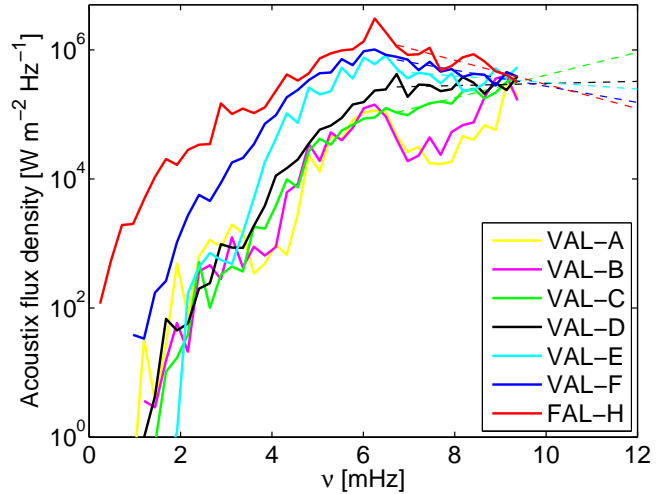}
\caption{Frequency distribution of the deposited acoustic flux averaged for individual model atmospheres. Dashed lines show linear extrapolations to higher frequencies.}
       \label{Fig:acfdens}
   \end{figure}
%

It is known that frequencies higher than our detectable limit $\nu_{\rm max}=9.6$~mHz are important for chromospheric heating \citep[e.g.,][]{Fossum06,Bello09,Bello10}. We calculated frequency-dependent densities of the deposited acoustic flux and averaged them over areas assigned to the models VAL A--F and FAL H (Fig.~\ref{Fig:acfdens}). We can see in the figure that the contribution of high frequencies is indeed significant. It is also seen that the low-frequency ($\nu < \nu_{\rm max}$) contribution increases in areas with enhanced activity, particularly in bright network elements (models VAL E, F) and plages (FAL~H), while a linear extrapolation indicates that the high-frequency contribution probably decreases in these areas.

%
   \begin{figure}
       \centering
\includegraphics[width=0.45\textwidth]{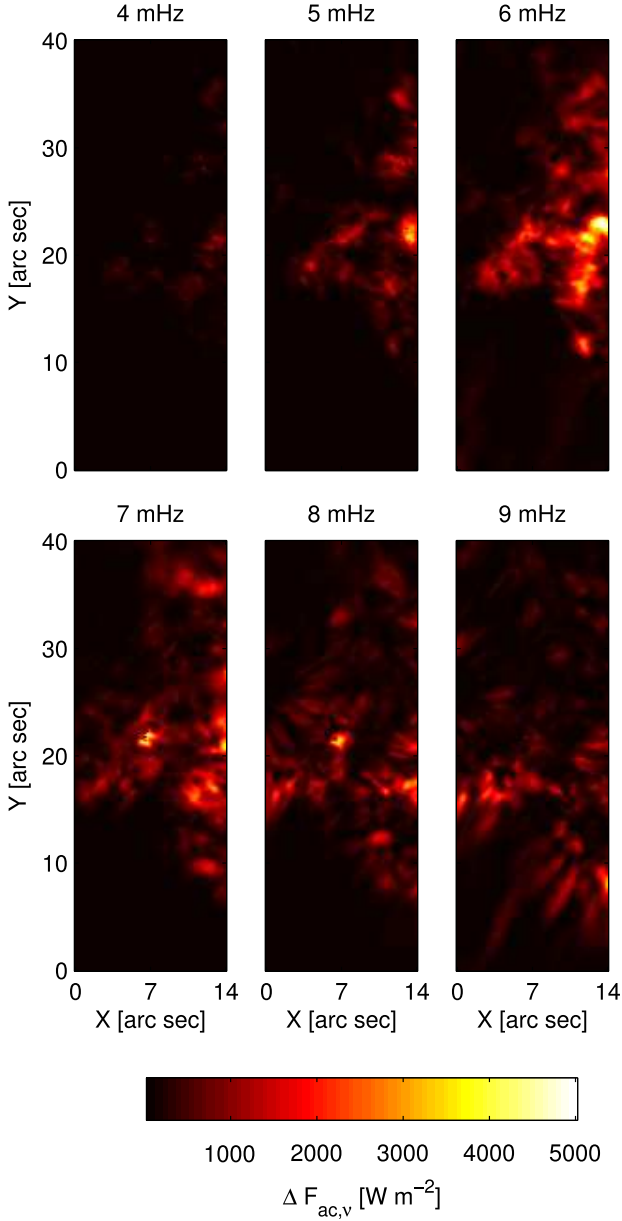}
\caption{Maps of deposited acoustic flux in six frequency bands from 4 mHz to 9 mHz.}
       \label{Fig:acfbands}
   \end{figure}
%

Maps of deposited acoustic energy flux in six frequency bands (FWHM = 0.86 mHz) at 4--9~mHz are displayed in Fig.~\ref{Fig:acfbands}. The fluxes in the bands at 1--3 mHz are small, with values less than 500~W m$^{-2}$. We can see in the maps that the deposited flux is concentrated in the plage region at frequencies up to 6 mHz and it is more dispersed, weaker in the plage and stronger in quiet chromosphere at higher frequencies. This supports the idea that low-frequency waves are important for the heating of active chromosphere, while high-frequency waves have a more significant contribution in quiet regions.

\section{Comparison of the deposited acoustic flux to radiative losses}
\label{Sect:6}

%
   \begin{figure}
       \centering
\includegraphics[width=0.48\textwidth]{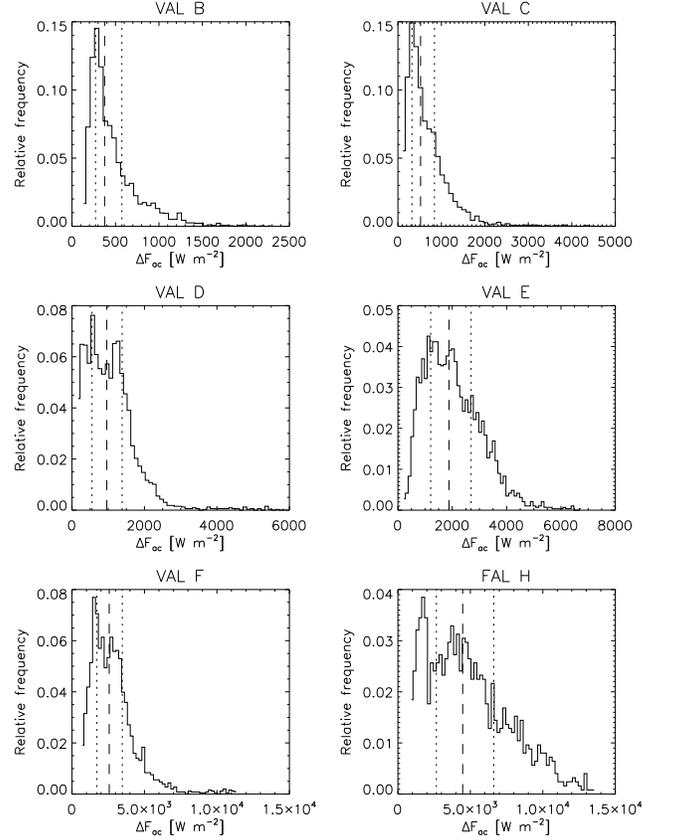}
\caption{Histograms of deposited acoustic flux for areas occupied by models VAL~B--F and FAL~H. Vertical dashed lines represent median values and dotted lines median deviations.}
       \label{Fig:hist}
   \end{figure}
%

It is seen from Figs.~\ref{Fig:lossmap} and \ref{Fig:acflux} that the deposited acoustic flux is indeed enhanced in the plage region. The coefficient of spatial correlation between the radiative losses and deposited acoustic flux is 0.72. However, there are many areas with high radiative losses where the deposited acoustic flux energy is quite small and vice versa. Therefore, we have to compare both quantities statistically. Figure~\ref{Fig:hist} shows histograms of deposited acoustic fluxes found in areas assigned to the models VAL B--F and FAL H. The histograms are clearly asymmetric, so that the statistics of acoustic fluxes is best represented by median values. The scatter of asymmetrically distributed data is characterized by negative and positive median deviations, $MND$ and $MPD$, defined separately for the data smaller and larger than the median, respectively:
\begin{eqnarray}
 MND = {\rm median}[X - {\rm median}(X)]_{<0} \nonumber \\
 MPD = {\rm median}[X - {\rm median}(X)]_{>0}  ,
\label{eq:md}
\end{eqnarray}
where $X$ stands for the data set. This definition is similar to that of median absolute deviation $MAD$, where absolute differences between the data and median are taken. For the normal distribution, $MAD \simeq 0.674 \ \sigma$ \citep{Rousse93}. Median values of the deposited acoustic flux in areas assigned to the individual models, together with median deviations, are listed in Table~\ref{Tab:acflux}. Contributions of median acoustic energy fluxes to the total integrated radiative losses are shown in the last column of the table. This contribution is small (0.15) in quiet Sun areas represented by models VAL A--C and increases gradually in active areas (models VAL D--F), reaching a maximum of 0.54 in the plage (model FAL~H). We remind that the maximum detected wave frequency is 9.6~mHz.

%
\begin{table}
\caption{Median deposited acoustic flux and its share in radiative losses.}
\label{Tab:acflux}
\centering
\begin{tabular}{ccccc}
\hline\hline
\noalign{\smallskip}
    Model & Median [W m$^{-2}$] & $MND$ & $MPD$ & $L$ share \\
\hline
\noalign{\smallskip}
    VAL A &   290 & $-80$  & 110 & 0.16 \\
    VAL B &   370 & $-100$ & 200 & 0.14 \\
    VAL C &   520 & $-200$ & 320 & 0.14 \\
    VAL D &   960 & $-410$ & 420 & 0.23 \\
    VAL E &  1880 & $-670$ & 820 & 0.37 \\
    VAL F &  2570 & $-850$ & 900 & 0.43 \\
    FAL H &  4470 & $-1820$ & 2150 & 0.54 \\
\hline
\end{tabular}
\end{table}
%

%
   \begin{figure}
       \centering
\includegraphics[width=0.48\textwidth]{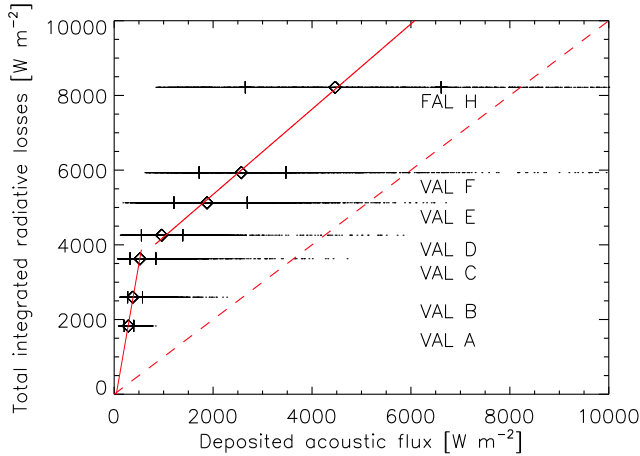}
\caption{Total radiative losses versus deposited acoustic flux. Median values (diamonds) and median deviations of the acoustic flux are shown together with linear fits for quiet and active areas. The dashed diagonal line represents the full balance of radiative losses by acoustic flux deposit.}
       \label{Fig:srov}
   \end{figure}
%

The total integrated radiative losses and the deposited acoustic flux are compared in a scatter plot shown in Fig.~\ref{Fig:srov}, including the median values and median deviations of the acoustic flux. Despite of a quite rough sampling by only seven model atmospheres, the difference between the acoustic energy input in quiet (non-magnetic) and active (magnetic) areas is clearly seen. It is interesting that the relation between the radiative losses $L$ and the deposited acoustic flux can be expressed by two separate linear fits
\begin{eqnarray}
L_{\rm q} = -300 + 7.63 \ \Delta F_{\rm ac} \nonumber \\
L_{\rm a} = 3100 + 1.14 \ \Delta F_{\rm ac}
\label{eq:linfits}
\end{eqnarray}
for quiet and active areas, respectively. We also see from the figure that at some sporadic locations in the active areas, the acoustic flux fully balances the radiative losses. Note that the parameters of the fit do not have a clear physical meaning due to the large scatter of acoustic fluxes and the uncertainties mentioned in Sect.~\ref{Sect:5}. However, we may speculate that the high value of the slope in quiet areas corresponds to a small share of acoustic energy flux in chromospheric heating, while the slope near to one in active areas indicates a more important contribution. We could also speculate that the constant term of 3100~W m$^{-2}$ might be related to the contribution of high-frequency waves (see Sect.~\ref{Sect:7}) or to other heating mechanisms.

\section{Discussion and conclusions}
\label{Sect:7}

We make, for the first time to our knowledge, a quantitative observational comparison of energy flux deposited by (magneto)acoustic waves with radiative losses both in the active and quiet chromospheres. The comparison is based on a consistent set of high-resolution observations made by the {\em IBIS} instrument. The studied region includes a weak plage with a maximum magnetic field strength of 1300 G and a quiet area. We focus on the middle chromosphere, where the core of the observed \ion{Ca}{2} 854.2~nm line is formed. The deposited acoustic flux is derived from Doppler velocities observed in the line core and the calculation of radiative losses is based on a grid of seven 1D semi-empirical models assigned to different parts of the region according to the best fit of synthetic \ion{Ca}{2} profiles to the observed ones averaged in time.

The radiative losses and deposited acoustic flux are spatially correlated (correlation coefficient 0.72). This supports the importance of acoustic waves for chromospheric heating. We show that in quiet region, the contribution of acoustic energy flux to radiative losses is small, only of about 15~\%. This can be explained by the facts that acoustic waves with frequencies below the photospheric cutoff $\nu_{\rm c} = 5.2$~mHz are trapped in the photosphere and that our measurements are limited to the maximum frequency of 9.6 mHz. 
Another situation is in active areas with photospheric magnetic field strength between 300 G and 1300 G and inclination of 20\degr--60\degr, according to our measurements. The contribution of acoustic energy flux to radiative losses increases from 23~\% (model VAL~D, chromospheric network) to 54~\% (model FAL~H, a plage). The coefficient of spatial correlation between the radiative losses and magnetic field inclination is 0.63. These facts clearly support the idea of {\em magnetic portals} \citep{Jefferies06}, where low-frequency magneto-acoustic waves can propagate into the chromosphere.

In summary, our estimate of the contribution of acoustic energy flux to the heating of magnetic areas and plages is significant but still insufficient to balance the radiative cooling. However, as we mentioned in Sect.~\ref{Sect:5}, our estimate of the deposited acoustic flux is only a lower limit. This is mainly due to the missing information about waves with frequencies higher than 9.6~mHz and the unknown transfer function $T\!F(\nu)$ in the chromosphere that we set to one. \citet{Fossum06} and \citet{Bello09,Bello10} studied the propagation of acoustic waves in quiet regions and from their results it follows that the relative contribution of waves with frequencies larger than 10 mHz is of about 50--100~\% of the acoustic flux measured in the range 5--10 mHz. On the other hand, our results indicate that in active regions, this relative contribution probably decreases with increasing activity, from model VAL~D to FAL~H (Sect.~\ref{Sect:5}, Fig.~\ref{Fig:acfdens}).

We can speculate that when the high-frequency waves would be included in the deposited acoustic flux estimation, the slope in the linear fit (Equation~7) would decrease by the factor of 0.5--0.7 in quiet areas, which means that the acoustic energy flux would still be smaller than the radiative losses (cf. Fig.~\ref{Fig:srov}).  In active areas, the fit would change in both parameters: The slope will not decrease as much as in the quiet areas (or it may even increase) and thus the constant term 3100~W~m$^{-2}$ would decrease. Then it cannot be excluded that the acoustic energy flux can approach to the radiative losses and may become the main contributor to the heating of bright network elements and plages. Observations with temporal resolution of at least 10~s are necessary to confirm that.

Our results, based on 1D hydrostatic semi-empirical models and a statistical approximation of estimated acoustic energy fluxes do not take into account the highly dynamic nature of energy-transfer processes in the solar chromosphere on time scales of observed wave periods. However, they may provide a useful general estimate of the role played by acoustic waves in the heating of long-lived chromospheric structures.

\acknowledgments
This work was supported by the Czech Science Foundation under the grant 14-04338S and by the project RVO:67985815 of the Academy of Sciences of the Czech Republic.
The Dunn Solar Telescope is located at the National Solar Observatory (NSO), which is operated by the Association of Universities for Research in Astronomy, Inc. (AURA), for the National Science Foundation.
{\em IBIS} has been built by INAF/Osservatorio Astrofisico di Arcetri with contributions from the Universities of Firenze and Roma ``Tor Vergata'', NSO, and the Italian Ministries of Research (MIUR) and Foreign Affairs (MAE). The observations at {\em IBIS} have been supported by the Rome Tor Vergata ``Innovative techniques and technologies for the study of the solar magnetism'' grant funded by MIUR.”

\bibliographystyle{aasjournal}
\bibliography{ArXiV_AAS00257}

\end{document}